\documentclass[twocolumn]{autart}

\usepackage{amsmath,amsfonts,amssymb}
\usepackage{graphicx}      
\usepackage{graphics}

\newcommand{\be}{\begin{equation}}
\newcommand{\ee}{\end{equation}}
\newcommand{\bd}{\begin{displaymath}}
\newcommand{\ed}{\end{displaymath}}
\newcommand{\bea}{\begin{eqnarray}}
\newcommand{\eea}{\end{eqnarray}}

\newcommand{\R}{\mathbb{R}}
\newcommand{\C}{\mathbb{C}}

\newcommand{\Rp}{\mathbb{R}_0^{+}}
\newcommand{\psa}{pseudospectral abscissa{ }}
\newcommand{\ps}{pseudospectrum }
\newcommand{\w}{\omega}
\newcommand{\alpf}{\alpha_f^N(\sigma)}

\newcommand{\In}{I_{n}}

\newtheorem{theorem}{Theorem}[section]
\newtheorem{definition}[theorem]{Definition}
\newtheorem{lemma}[theorem]{Lemma}
\newtheorem{proposition}[theorem]{Proposition}

\begin{document}
\begin{frontmatter}

\title{A Predictor-Corrector Type Algorithm for the Pseudospectral Abscissa Computation of Time-Delay Systems}

\author[First]{Suat Gumussoy},
\author[First]{Wim Michiels}

\address[First]{Department of Computer Science, K. U. Leuven, \\
        Celestijnenlaan 200A, 3001, Heverlee, Belgium \\
        \mbox{(e-mail: suat.gumussoy@cs.kuleuven.be, wim.michiels@cs.kuleuven.be)}.}


\begin{abstract}
The pseudospectrum of a linear time-invariant system is
the set in the complex plane consisting of all the roots
of the characteristic equation when the system matrices
are subjected to all possible perturbations with a given
upper bound. The pseudospectral abscissa is defined as
the maximum real part of the characteristic roots in the
pseudospectrum and, therefore, it is for instance
important from a robust stability point of view. In this
paper we present an accurate method for the computation
of the pseudospectral abscissa of retarded delay
differential equations with discrete pointwise delays.
Our approach is based on the connections between the
pseudospectrum and the level sets of an appropriately
defined complex function. The computation is done in two steps. In the prediction step, an approximation of the \psa is obtained based on a rational approximation of the characteristic matrix and the application of a bisection algorithm. Each step in this bisection algorithm relies on checking the presence of the imaginary axis eigenvalues of a complex matrix, similar to the delay free case. In the corrector step, the approximate pseudospectral
abscissa is corrected to any given accuracy, by solving a
set of nonlinear equations that characterize extreme
points in the pseudospectrum contours.
\end{abstract}

\end{frontmatter}

\section{Introduction}
The pseudospectrum provides information about the
characteristic roots of a system when the system
matrices in the characteristic equation are subject to
perturbations. It is closely related to the robust
stability of a system and to the distance to instability,
\cite{Trefethen:97}. We consider the time-delay
system
\be \label{eq:DDEF}
\dot{x}(t)=\sum_{i=0}^m
A_i x(t-\tau_i),
\ee
where $A_i\in\R^{n\times n}$, $\tau_0=0$
$\tau_i\in\Rp$ for $i=1,\ldots,m$ and define $\tau_{\max}$ as the maximum delay of the time-delay system,
\bd
\tau_{\max}:=\max\{\tau_0,\ldots,\tau_m\}.
\ed Note that this type of time-delay system is of retarded type \cite{WimBook}.

The characteristic equation of the time-delay system
(\ref{eq:DDEF}) is:
\be \label{eq:chareqn} \det
F(\lambda)=0 \ee where \be \label{eq:F}
F(\lambda):=\lambda I_n-\left(\sum_{i=0}^m
A_ie^{-\lambda\tau_i}\right).
\ee

The characteristic
equation (\ref{eq:chareqn}) has infinitely many roots
extending to the complex left half-plane, yet a finite
number of roots in any right half plane \cite{WimBook}. Therefore the
maximum of the real parts of the characteristic roots is
well defined, and called the \emph{spectral abscissa}
\be \label{eq:sa}
\alpha(F):=\max_{\lambda\in\C}\{\Re(\lambda): \det
F(\lambda)=0 \}.
\ee

The $\epsilon$-\emph{pseudospectrum} of the function $F$
is the collection of characteristic roots of
(\ref{eq:DDEF}) when the system matrices are subject to
all possible perturbations with a given upper bound
determined by $\epsilon>0$ and individual weights on the system matrices. More precisely, it is defined
as
{\small
\begin{multline} \label{eq:psF1}
\Lambda_\epsilon(F):=\left\{\lambda\in\C: \det\left(\lambda I_n-\left(\sum_{i=0}^m (A_i+\delta A_i) e^{-\lambda\tau_i}\right)  \right)=0 \right.\\
\text{ for some } (\delta A_0,\ldots,\delta
A_m)\in\C^{n\times n\times(m+1)}
 \\ \left. \text{ satisfying }\sigma_{\max}(\delta A_i)\leq\frac{\epsilon}{w_i} \text{ for }
\mbox{i=0,\ldots,m} \right\}.
\end{multline}
}

\noindent Here the numbers $w_i\in\Rp\cup\{\infty\},$
\mbox{$i=0,\ldots,m$}, are weights on the perturbations
of the system matrices $A_i$  which can be chosen a
priori. A weight equal to infinity means that no
perturbations on the corresponding matrix are assumed. Note the $\epsilon$-\emph{pseudospectrum} of the function $F$ depends on $\epsilon$ and the chosen weights on system matrices $w_i$ for $i=0,\ldots,m$.

 The maximum real part in the \ps is the \emph{\psa} which is defined as
\be \label{eq:psa1}
\alpha_\epsilon(F)=\sup_{\lambda\in\C}\{\Re(\lambda):
\lambda\in\Lambda_\epsilon(F) \}. \ee

The \psa is a bound characterizing the stability robustness of the system. All characteristic roots of the time-delay system (\ref{eq:chareqn}) are on the left complex half-plane for all possible perturbations as in (\ref{eq:psF1}) if and only if $\alpha_\epsilon<0$, therefore, the system (\ref{eq:DDEF}) is robustly stable. Similarly, the inequality \mbox{$\alpha_\epsilon<-\sigma_0$} (where $\sigma_0>0$) is a necessary and sufficient condition guaranteeing that all characteristic roots lie to the left of $\Re(s) = -\sigma_0$. This type of stability is known as $\Gamma$-stability in the literature where the $\Gamma$-region is the half-plane $\Re(s)<\sigma_0$ and it gives an upper bound for the exponential  rate of convergence of a system. Note that there are many sufficient conditions to check robust stability or $\Gamma$-stability in the presence of perturbations at system matrices in the literature, for instance, conditions based on Lyapunov functional approach as in  \cite{ShuZhan:09}, \cite{Kharitonov:06} or conditions based on matrix measures as in \cite{Cao:03}, \cite{Wang:98}.

In the finite-dimensional, delay-free case, (\ref{eq:F}) reduces to
\be \label{bla2}
 F_0(\lambda)=\lambda I_n-A_0, \ee and the \ps (for a
unity weight) can be equivalently expressed as
 \be \label{bla}
\Lambda_\epsilon(F_0)=\left\{\lambda\in\C:
\sigma_{\max}\left(F_0(\lambda)^{-1}\right)
>\frac{1}{\epsilon}\right\},\ee (see \cite{Boyd:89}).
Thus, the boundaries of the \ps  can be computed as the
level set of a resolvent norm. This connection is used to
compute the distance to instability and the \psa via a
bisection algorithm in \cite{Byers:88} and
\cite{Burke:03} respectively. A quadratically convergent
algorithm for the \psa computation is given in
\cite{BurkeCC:03}, based on a `criss-cross' procedure.

In \cite{Michiels:06} the formula (\ref{bla}) is
generalized from (\ref{bla2}) to a broad class of matrix
functions including (\ref{eq:F}). In particular, from
Theorem $1$ of \cite{Michiels:06} it follows that the
$\epsilon$-pseudospectrum of (\ref{eq:F}), as defined by
(\ref{eq:psF1}), can be equivalently expressed as \be
\label{eq:psF2} \Lambda_\epsilon(F)=\left\{\lambda\in\C:
f(\lambda)
>\frac{1}{\epsilon}\right\} \ee where \be \label{eq:f}
f(\lambda)=w(\lambda)\sigma_{\max}(F(\lambda)^{-1}) ,\ \
w(\lambda)=\sum_{i=0}^m
\frac{e^{-\Re(\lambda)\tau_i}}{w_i}. \ee

Using the formula (\ref{eq:psF2}), the \psa
(\ref{eq:psa1}) can be rewritten as \be \label{eq:psa2}
\alpha_\epsilon(F)=\max_{\lambda\in\C}\left\{\Re(\lambda):
f(\lambda)=\frac{1}{\epsilon} \right\}. \ee Note that the
maximum in (\ref{eq:psa2}) is well-defined since
$F(\lambda)^{-1}$ is a strictly proper function and
$w(\lambda)$ is uniformly bounded on any complex right
half-plane.

Our main contribution is the extension of the pseudospectral abscissa computation to infinite-dimensional time-delay systems. Both in the definition of the pseudospectrum and in the computational scheme the structure of the delay equation is fully exploited. The numerical methods in \cite{Burke:03}, \cite{BurkeCC:03} consider the finite-dimensional, delay-free case. Our algorithm for the \psa computation of time-delay
systems is implemented in two steps: a prediction and a
correction step. First the transcendental function
(\ref{eq:F}) is approximated by a rational function in
Section~\ref{sec:approx}, and an approximation of the
\psa is computed using this rational approximation in
Section~\ref{sec:psapred}. Second, in
Section~\ref{sec:psacorr} the approximate result is
corrected  using a locally convergent method which is based on solving equations characterizing extreme values in the pseudospectrum contour. The overall algorithm
for the \psa computation is outlined in Section
\ref{sec:alg}. A numerical example and concluding remarks
can be found in Sections~\ref{sec:ex} and \ref{sec:conc}.

\textbf{Notation:} \\
The notations in the paper are standard and given below.
\begin{tabbing}
  \= ${\textstyle \sigma_{\max}(A)}$ \=: the largest singular value of the matrix $A$ \\
  \> $A^{*}$ \>: complex conjugate transpose of the matrix $A$ \\
  \> $I_n$ \>: identity matrix with dimensions $n \times n$ \\
  \> $0_n$ \>: zero matrix with dimension $n \times n$ \\

  \> $\C, \R$ \>: the field of the complex and real numbers \\
  \> $\Rp$  \> : the positive real numbers, excluding zero \\
  \> $\Re(u)$ \> : real part of the complex number $u$ \\
  \> $\Im(u)$ \> : imaginary part of the complex number $u$ \\
  \> $|u|$ \> : magnitude of the complex number $u$ \\
  \> $\bar{u}$ \> : conjugate of the complex number $u$ \\
  \> $\mathcal{D}(.)$ \> : domain of an operator \\
  \> $\mathcal{C}, \mathcal{L}_2$ \> : the space of continuous and square integrable \\
  \> \> $\ \ $complex functions, i.e., $\mathcal{L}_2([-\tau_{\max},0],\C^n):=$ \\
  \> \> $\ \ $ $\{f:[-\tau_{\max},0] \rightarrow \C^n : \int_{-\tau_{\max}}^0 |f(t)|^2 dt <\infty\}$ \\
  \> $\|F\|_\infty$ \> : $\mathcal{L}_\infty$ norm of the transfer function $F(j\omega)$ \\
  \> $\alpha(G)$ \>: the spectral abscissa of $G$, i.e.,\\
  \> \> $\ \ $$\sup_{\lambda\in\C}\left\{\Re(\lambda): \det(G(\lambda))=0 \right\}$.
\end{tabbing}

\section{FINITE DIMENSIONAL APPROXIMATION} \label{sec:approx}

We derive a  rational approximation of the function
$F(\lambda)$, given by $(\ref{eq:F})$, which is
instrumental to the algorithm developed in the next
sections. It is based on a finite-dimensional
approximation of the system
 \bea
\label{eq:DDEFyu} \dot{x}(t)&=&\sum_{i=0}^m A_i
x(t-\tau_i)+u(t),\ y(t)=x(t), \eea
whose input-output map is characterized by the transfer
function $F(\lambda)^{-1}$.

We start by reformulating the system (\ref{eq:DDEFyu}) as
an infinite-dimensional linear system in the standard
form, \cite{Curtain:95}. When defining the space $X:=\C^n\times
\mathcal{L}_2([-\tau_{\max},0],\C^n)$ equipped with the
inner product \bd
<(y_0,y_1),(z_0,z_1)>_X=<y_0,z_0>_{\C^n}+<y_1,z_1>_{\mathcal{L}_2},
\ed we can rewrite (\ref{eq:DDEFyu}) as
 \bea \label{eq:ODEA}
\dot{z}(t)&=&\mathcal{A}z(t)+\mathcal{B}u(t),\\
\nonumber y(t)&=&\mathcal{C}z(t),
\eea where
\begin{multline} 
 \mathcal{D}(\mathcal{A}) =\{z=(z_0,z_1)\in X: z_1
\textrm{ is absolutely continuous} \\
 \textrm{on } [-\tau_{\max},0], \frac{dz_1}{d\theta}\in\mathcal{C}([-\tau_{\max},0],\C^n), z_0=z_1(0)\}, \\
\end{multline}
\bea \nonumber \mathcal{A}z&=&\left(\begin{array}{c}
                     A_{0}z_0+ \sum_{i=1}^m A_i z_1(-\tau_i)\\
                     \frac{dz_1}{d\theta}(.)
                   \end{array}\right), z\in\mathcal{D}(\mathcal{A}),  \\
\nonumber \mathcal{B}u&=&\left(\begin{array}{c}
                     u\\
                     0
                   \end{array}\right), u\in\C^n,\  \ \mathcal{C}z=z_0,\  z\in X. \eea The connection between (\ref{eq:DDEFyu})
and (\ref{eq:ODEA}) is that $z_0(t)\equiv x(t)$,
$z_1(t)\equiv x(t+\theta), \theta\in[-\tau_{\max},0]$.

Next, we discretize the infinite-dimensional system
(\ref{eq:ODEA}). We use a spectral method, as in
\cite{Breda:05,Breda:06}. Given a positive integer $N$,
we consider a mesh $\Omega_N$ of $N+1$ distinct points in
the interval $[-\tau_{\max},\ 0]$,
\begin{equation}\label{defmesh}
\Omega_N=\left\{\theta_{N,i},\ i=-N,\ldots,0\right\},
\end{equation}
where we assume that $\theta_{N,0}=0$. With the Lagrange
polynomials $l_{N,k}$ defined as real valued polynomials
of degree $N$ satisfying
\[
l_{N,k}(\theta_{N,i})=\left\{\begin{array}{ll}1 & i=k\\
0 & i\neq k
\end{array}\right.
\]
where $i,k\in\{-N,\ldots,0\}$.
We can construct a $N+1$ by $N+1$ differentiation matrix
on the mesh $\Omega_N$,
\be \label{eq:D}
D:=\left[\begin{array}{ccc|c}
d_{-N,-N} & \cdots & d_{-N,-1} & d_{-N,0} \\
\vdots & & \vdots & \vdots \\
d_{-1,-N}& \cdots & d_{-1,-1}  & d_{-1,0}\\ \hline
d_{0,-N} & \cdots & d_{0,-1} & d_{0,0}
\end{array}\right]=\left[\begin{array}{c|c}
D_{1,1} & D_{1,2} \\ \hline D_{2,1} &
D_{2,2}\end{array}\right],
\ee
where
\begin{equation}\label{defD}
d_{i,k}=l^{\prime}_{N,k}(\theta_{N,i}) ,\ \ \ \
i,k\in\{-N,\ldots,0\}.
\end{equation}

Then, similarly as in \cite{Breda:05}, the delay
differential equation can be approximated by the
finite-dimensional system:
\begin{equation}\label{approx}
\begin{array}{l}
\dot z(t)= {\bf A_N} z(t)+{\bf B_N} u(t),\ z(t)\in\R^{(N+1)n\times 1} \\
y(t)={\bf B_N}^* z(t)
\end{array}
\end{equation}
where
\be \label{eq:AN}
\begin{array}{l}
{\bf A_N}=\left[\begin{array}{llll}
d_{-N,-N}\In &\hdots & d_{-N,-1}\In & d_{-N,0}\In \\
\vdots & &  \vdots & \vdots \\
d_{-1,-N}\In &\hdots & d_{-1,-1}\In  & d_{-1,0}\In \\
\Gamma_{-N} & \hdots & \Gamma_{-1} & \Gamma_0
\end{array}\right],
\end{array}
\ee
\[
\begin{array}{lll}
\Gamma_0&=& A_0+\sum_{l=1}^m A_l l_{N,0}(-\tau_l),\\
\Gamma_{k}&=&\sum_{l=1}^m A_l l_{N,k}(-\tau_l),\ \ \
k\in\{-N,\ldots,-1\}, \\
{\bf B_N}&=&[0_n\ \ldots\ 0_n\ I_n]^*.
\end{array}
\]

In order to explain the effects of the approximation of
(\ref{eq:DDEFyu}) by (\ref{approx}) in the frequency
domain, we need the following definition.
\begin{definition}\label{Defpn}
For $\lambda\in\mathbb{C}$, let $p_N(\cdot;\ \lambda)$ be
the polynomial of degree $N$ satisfying
\begin{equation}\label{defpn}
\begin{array}{l}
p_N(0;\ \lambda)=1,\\
p_N^{\prime}(\theta_{N,i};\ \lambda)=\lambda
p_N(\theta_{N,i};\ \lambda),\ \ i\in\{-N,\ldots,-1\}.
\end{array}
\end{equation}
\end{definition}

Note that the polynomial $p_N(t;\ \lambda)$ is an
approximation of $e^{\lambda t}$ on the interval $[-1;\
0]$. Indeed, the first equation of (\ref{defpn}) is an
interpolation requirement at zero, the other equations
are collocation conditions for the differential equation
$\dot z=\lambda z$, of which $e^{\lambda t}$ is a
solution.

We can now state the main result of this section:
%
\begin{theorem} \label{thm:ANGN} The transfer function of the
system (\ref{approx}) is given by
\begin{multline}\label{eq:FNpn} {\bf B_N}^*(\lambda
I_{(N+1)n}-{\bf A_N})^{-1}{\bf B_N}\\
= \left(\lambda I_n-A_0-\sum_{i=1}^m A_i p_N(-\tau_i;\
\lambda)\right)^{-1},
\end{multline}
where the function $p_N$ is given by
Definition~\ref{Defpn}. \label{maintheo}
\end{theorem}

For the proof of the theorem we refer to Section
\ref{sec:app} of the appendix.

\medskip

Recall that the transfer function of (\ref{eq:DDEFyu}) is
given by $F(\lambda)^{-1}$. Therefore, the effect of
approximating (\ref{eq:DDEFyu}) by the finite-dimensional
system (\ref{approx}) can be interpreted as the effect of
approximating the function $F(\lambda)$ by \be
\label{eq:FN} F_N(\lambda):=\lambda I_n-A_0-\sum_{i=1}^m
A_ip_N(-\tau_i,\lambda). \ee
In Proposition~\ref{lem1} of the appendix it is shown
that the functions
\[
\lambda\mapsto p_i(-\tau_i;\ \lambda)
\]
are proper rational functions. Hence, the function
$F_N(\lambda)$ can be considered as a rational
approximation of $F(\lambda)$.

\noindent \textbf{Remark:} It follows from Theorem~\ref{thm:ANGN} that
\bd
\alpha(F_N)=\sup_{\lambda\in\C} \{\Re(\lambda): \det(\lambda I_{(N+1)n}-{\bf A_N})=0 \}.
\ed

\section{Approximation of the Pseudospectral Abscissa} \label{sec:psapred}

Given the approximation (\ref{eq:FN}) of $F(\lambda)$ and
the characterization  (\ref{eq:psa2}) we can obtain an
approximation of the pseudospectral abscissa
$\alpha_{\epsilon}(F)$ by computing
 \be\label{eq:psaN}
\alpha_\epsilon^N(F):=\max_{\lambda\in\C}\left\{\Re(\lambda):
f_N(\lambda)=\frac{1}{\epsilon} \right\}, \ee where \be
\label{eq:fN}
f_N(\lambda)=w(\lambda)\sigma_{\max}(F_N(\lambda)^{-1})
,\ \ w(\lambda)=\sum_{i=0}^m
\frac{e^{-\Re(\lambda)\tau_i}}{w_i}. \ee This is outlined
in what follows.

Let the function $\alpha_f^N$ be defined on the interval $(\alpha(F_N),\ \infty)$ by \be \label{eq:alpfN}
\alpha_f^N(\sigma)=\sup_{\w\in\R}f_N(\sigma+j\w). \ee


\begin{prop} \label{prop:alpfmonot}The function $\alpha_f^N$ has
the following properties.
\begin{enumerate}
\item It is strictly
decreasing.
\item $\lim_{\sigma\rightarrow\alpha(F_N){+}} \alpf = +\infty.$
\item $\lim_{\sigma\rightarrow +\infty} \alpf = 0.$
\item $\alpha_{\epsilon}^N(F)=\left\{\sigma\in(\alpha(F_N),\ \infty):\ \alpha_f^N(\sigma)=\frac{1}{\epsilon}\right\}.$
\end{enumerate}
\end{prop}
\noindent\textbf{Proof.\ } We have
\[
\alpha_f^N(\sigma)=w(\sigma)
\sup_{\omega\in\mathrm{R}}\sigma_{\max}\left(F_N(\sigma+j\omega)^{-1}\right).
\]
For the first assertion, note that the function
$\sigma\mapsto w(\sigma)$ is strictly decreasing.
Furthermore, the function
\[
\sigma\mapsto\sup_{\omega\in\mathrm{R}}\sigma_{\max}\left(F_N(\sigma+j\omega)^{-1}\right)
\]
cannot be increasing because this would be in
contradiction with the fact that the sets
\[
\left\{\lambda\in\mathbb{C}:\
\sigma_{\max}(F_N(\lambda)^{-1})>\frac{1}{\epsilon}\right\}
\]
can be interpreted as pseudospectrum of the function
$F_N$, where only $A_0$ is perturbed (see \cite{Michiels:06} for
the details).

%
%

 The second assertion follows from the
fact that $F_N$ has a zero on the boundary
$\Re(\lambda)=\alpha(F_N)$.
The third assertion is due to the fact that $F_N^{-1}$ is
strictly proper. The last assertion follows from the
other assertions.\hfill $\Box$

Proposition~\ref{prop:alpfmonot} directly leads to a
bisection algorithm over the interval $(\alpha(F_N),\
\infty)$ for the computation of $\alpha_{\epsilon}^N(F)$,
where the main step consists of checking whether or not
the inequality
 \be
\label{checkcond} \alpha_f^N(\sigma)>\frac{1}{\epsilon}
\ee
is satisfied. Using Theorem~\ref{maintheo}, we get
\begin{multline}
\nonumber \label{ineq:AN} \alpha_f^N(\sigma)>\frac{1}{\epsilon}
\Leftrightarrow
w(\sigma)\sup_{\omega\in\mathbb{R}}\sigma_{\max}\left(F_N(\sigma+j\omega)^{-1}\right)>\frac{1}{\epsilon}\Leftrightarrow
\\
 \sup_{\omega\in\mathbb{R}}\sigma_{\max}\left({\bf B_N}^*\left(j\w I_{(N+1)n}-\left({\bf A_N}-\sigma
I_{n(N+1)}\right)\right)^{-1}{\bf B_N}\right)\\
>\frac{1}{\epsilon w(\sigma)}.
\end{multline}

It follows that the inequality (\ref{checkcond}) is
satisfied if and only if the matrix \be\label{sna}
{\bf B_N}^*\left(j\w I_{(N+1)n}-\left({\bf A_N}-\sigma
I_{n(N+1)}\right)\right)^{-1}{\bf B_N} \ee
has a singular value
equal to $\frac{1}{w(\sigma)\epsilon}$ for some value of
$\omega$. According to~\cite{Byers:88}, this is
equivalent to requiring that the Hamiltonian matrix \be
\label{eq:HamAN} H_{N,\sigma}:=\left[
  \begin{array}{cc}
    {\bf A_N}-\sigma I_{n(N+1)} & (w(\sigma)\epsilon){\bf B_N B_N}^* \\
    -(w(\sigma)\epsilon){\bf B_N B_N}^* & -\left(({\bf A_N}-\sigma I_{n(N+1)}\right)^* \\
  \end{array}
\right] \ee has imaginary axis eigenvalues\footnote{These
are given by $j\omega$, where $\omega$ is such that the
matrix (\ref{sna}) has a singular value equal to
$(\epsilon w(\sigma))^{-1}$. }.

Putting together the above results we arrive at the
following algorithm for computing $\alpha_{\epsilon}^N(F)$, the approximation
 of $\alpha_{\epsilon}(F)$.
\begin{alg} \label{alg:bisect} \ \
${}$\\
\textit{Input:} system data, tolerance for the prediction step, tol, and number of
discretization points, $N$\\
\textit{Output:} the approximate pseudospectral abscissa, $\alpha_\epsilon^N(F)$, and the corresponding frequencies, $j\tilde \omega_i$
\begin{enumerate}
  \item[1)] $\sigma_L=\alpha(F_N)$, $\sigma_R=\infty$, $\Delta \sigma=$tol,
  \item[2)] while $(\sigma_R-\sigma_L)>\textrm{tol}$
  \begin{enumerate}
  \item[2.1)] if $(\sigma_R=\infty)$ \\
              then $\Delta \sigma = 2 \times \Delta \sigma$, $\sigma_M=\sigma_L+\Delta \sigma$, \\
              else $\sigma_M=\frac{\sigma_L+\sigma_R}{2}$.
  \item[2.2)] if $H_{N,\sigma_M}$ has imaginary axis eigenvalues  \\
              then $\sigma_L=\sigma_M$, \\
              else $\sigma_R=\sigma_M$.
  \end{enumerate}
   \item[] \hspace*{-0.8cm}\{result:
   $\alpha_{\epsilon}^N(F)=\sigma_L$,\
   $j\tilde \omega_i$:  imaginary axis eigenvalues of \\
   \hspace*{-0.7cm}$H_{N,{\sigma}_L}$\}
\end{enumerate}
\end{alg}
It is important to note that the algorithm does not
require an explicit computation of the rational function
$F_N$. This is due to Theorem~\ref{maintheo}.

\section{Correcting the pseudospectral abscissa} \label{sec:psacorr}

Algorithm \ref{alg:bisect} finds the complex points
\[
\tilde{\lambda}_i=\alpha_\epsilon^N(F)+j\tilde{\w}_i,\
i=1,\ldots,\tilde{n},
\]
which are approximations of the rightmost elements of the
pseudospectrum $\Lambda_{\epsilon}(F)$, the accuracy
depending on the tolerance and the number of
discretization points, $N$. These approximations can be
corrected by solving a set of  equations inferred from a
nonlinear eigenvalue problem. This is detailed in what
follows.

The function $\alpha_f(\sigma)$ can be defined in a
similar way as the function $\alpf$ as \be
\label{eq:alpf}
\alpha_f(\sigma):=\sup_{\w\in\R}f(\sigma+j\w), \ee where
$\sigma\in(\alpha(F),\ \infty)$. Using the arguments as
spelled out in the proof of
Proposition~\ref{prop:alpfmonot} it can be shown that
\be\label{xyz} \alpha_f(\sigma)=\frac{1}{\epsilon}
\ee
if and only if $\sigma=\alpha_\epsilon(F)$.

Using the definition  (\ref{eq:f}) of $f(\lambda)$ ,
the equality (\ref{xyz}) can be written as
\begin{multline}
\sup_{\w\in\R}
\sigma_{\max}\left( \left( (\sigma+j\w)I_n-\sum_{i=0}^m A_i
e^{-(\sigma+j\w)\tau_i}\right)^{-1}\right) \cdot\\ w(\sigma)
 =\frac{1}{\epsilon},
\end{multline}
or, equivalently, \be
\|F_\sigma(j\w)^{-1}\|_\infty=\frac{1}{\epsilon
w(\sigma)}, \ee where \be F_\sigma(j\w)=j\w
I_n-A_{\sigma,0}-\sum_{i=1}^m A_{\sigma,i}e^{-j\tau_i\w}
\ee and \bea \label{eq:Ashift} A_{\sigma,0}=A_0-\sigma
I_n, \quad A_{\sigma,i}=A_ie^{-\tau_i\sigma},\
i=1,\ldots,m. \eea

Similarly the connection between a transfer function
and the spectrum of a corresponding Hamiltonian matrix in
the finite dimensional case, the following lemma
establishes connections between the singular value curves
of $F_\sigma(j\omega)^{-1}$ and the spectrum of a
nonlinear eigenvalue problem.
\begin{lemma}\label{thm:FandH}
Let $\xi>0$ and $\sigma\in (\alpha(F),\ \infty)$.  The
matrix $F_\sigma(j\omega)^{-1}$ has a singular value
equal to $\xi$ for some $\omega\geq 0$ if and only if
$\lambda=j\omega$ is a solution of the equation
\begin{equation}\label{nonlinear-eigenv}
\det H(\lambda,\sigma,\xi)=0,
\end{equation}
where
\begin{equation}\label{defh}
H(\lambda,\sigma,\xi):=\lambda I-M_{\sigma,0}-\sum_{i=1}^m \left(M_{\sigma,i}
e^{-\lambda\tau_i}+M_{\sigma,-i}e^{\lambda\tau_i}\right),
\end{equation}
with
\[
\begin{array}{l}
M_{\sigma,0}=\left[\begin{array}{cc}  A_{\sigma,0}& \xi^{-2}I_n\\
-I_n & -A_{\sigma,0}^*,
\end{array}\right],\\
M_{\sigma,i}=\left[\begin{array}{cc} A_{\sigma,i} &0\\0&0
\end{array}\right],\ \
M_{\sigma,-i}=\left[\begin{array}{cc} 0
&0\\0&-A_{\sigma,i}^*
\end{array}\right],\ \ 1\leq i\leq N,
\end{array}
\]
\end{lemma}

\noindent\textbf{Proof.\ } The proof is similar to the
proof of Proposition~22 in \cite{Genin:02}. For all
$\omega\in\mathbb{R}$, we have the relation
\begin{multline}\label{vier}
\det H(j\omega,\sigma,\xi)\det(-\xi^2I_n)= \\ \det
((F_\sigma^{-1}(j\omega))^{*}F_\sigma^{-1}(j\omega)-\xi^2
I)\\
\det\left(\left[\begin{array}{cc}F_\sigma(j\omega) &0\\
-I_n& -(F_\sigma(j\omega))^*
\end{array}\right]\right),
\end{multline}
because both left and right hand side can be interpreted
as expressions for the determinant of the 2-by-2 block
matrix
\[
\left[\begin{array}{cc|c} F_\sigma(j\omega) &0& I_n\\
-I_n&-(F_\sigma(j\omega))^*&0_n\\
\hline 0_n&I_n&-\xi^2I_n
\end{array}\right]
\]
using Schur complements. We get from (\ref{vier}):
\[
\det ((F_\sigma^{-1}(j\omega))^{*}F_\sigma^{-1}(j\omega)-\xi^2
I)=0\Leftrightarrow\det H(j\omega,\sigma,\xi)=0.
\]
This is equivalent to the assertion of the theorem.
 \hfill $\Box$

For a given value of $\xi$ and $\sigma$ the solutions of
(\ref{nonlinear-eigenv}) can be found by solving the
nonlinear eigenvalue problem
\begin{equation}
H(\lambda,\sigma,\xi)\ v=0,\ \ v\in\mathbb{C}^{2n},\
v\neq 0, \label{prob:HxiEig}
\end{equation}
which in general has an infinite number of solutions.

\begin{figure}
\begin{center}
\resizebox{9cm}{!}{\includegraphics{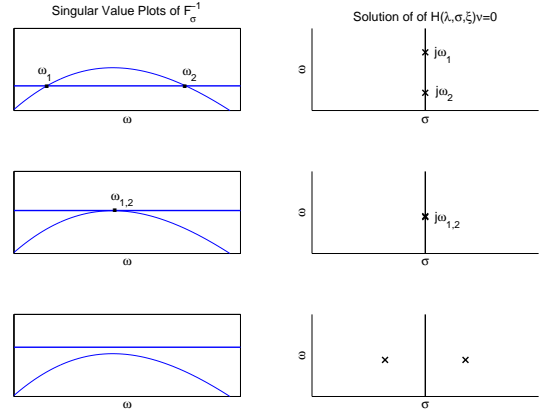}}
 \caption{\label{fig:semisimple} (left) Intersections of the
 singular value plot of $F_\sigma^{-1}$ with the horizontal line $\frac{1}{\epsilon w(\sigma)}$
 for the cases where (top) $\|F_\sigma(j\w)^{-1}\|_\infty>\frac{1}{\epsilon w(\sigma)}$, (middle) $\|F_\sigma^{-1}(j\w)\|_\infty=\frac{1}{\epsilon w(\sigma)}$ and
 (bottom) $\|F_\sigma^{-1}(j\w)\|_\infty<\frac{1}{\epsilon w(\sigma)}$. (right) Corresponding eigenvalues of the problem (\ref{prob:HxiEig})
 where $\xi=\frac{1}{\epsilon w(\sigma)}$.}
\end{center}
\end{figure}

The correction method is based on the property that if
$\sigma$ is such that
\[
\|F_{\sigma}(j\omega)^{-1}\|_\infty=\frac{1}{\epsilon\w(\sigma)},
\]
then the nonlinear eigenvalue problem
(\ref{prob:HxiEig}) has a multiple non-semisimple
eigenvalue for $\xi=\frac{1}{\epsilon w(\sigma)}$, as
clarified in Figure~\ref{fig:semisimple}.

Let $\alpha_\epsilon(F)+j\w_\epsilon$ be a rightmost
element of $\Lambda_{\epsilon}(F)$.
Setting
\[
h_{\sigma}(\lambda)=\det
H\left(\lambda,\sigma,\frac{1}{\epsilon
w(\sigma)}\right),
\]
the pair
$(\omega,\sigma)=(\w_\epsilon,\alpha_\epsilon(F))$
satisfies
\begin{equation}\label{direct1}
h_{\sigma}(j\omega)=0,\ \ h_{\sigma}^{\prime}(j\omega)=0.
\end{equation}
These complex-valued equations  seem over-determined but
this is not the case due to the spectral properties
of~$H$, which imply the following result.
\begin{prop}\label{corextra}
For $\omega\geq 0$, we have
\begin{equation}\label{col1}
\Im\ h_{\sigma}(j\omega)=0
\end{equation}
 and
\begin{equation}\label{col2}
 \Re\ h_{\sigma}^{\prime}(j\omega)=0.
\end{equation}
\end{prop}
\noindent\textbf{Proof.\ } It can easily be shown that
\[
h_{\sigma}(\lambda)=h_{\sigma}(-\lambda),\ \ \ h_{\sigma}^{\prime}(
\lambda)=-h_{\sigma}^{\prime}( -\lambda).
\]
Substituting $\lambda=j\omega$ yields
\[
\begin{array}{l}
h_{\sigma}(j\omega)=h_{\sigma}(-j\omega)=\left(h_{\sigma}(j\omega)\right)^*, \\
h_{\sigma}^{\prime}(j\omega)=-h_{\sigma}^{\prime}(-j\omega)=-\left(h_{\sigma}^{\prime}(j\omega)\right)^*,
\end{array}
\]
and the assertions follow. \hfill $\Box$

\medskip

\noindent Using Proposition~\ref{corextra} we can simplify the
conditions (\ref{direct1}) to:
\begin{equation}\label{direct2x}
\left\{\begin{array}{l}
\Re\ h_{\sigma}(j\omega)=0, \\
\Im\ h_{\sigma}^{\prime}(j\omega)=0.
\end{array}\right.
\end{equation}
Hence, the pair $(\w_\epsilon,\alpha_\epsilon(F))$ can be
directly computed by solving the two equations
(\ref{direct2x}) for $\omega$ and $\sigma$, e.g.\ using
Newton's method, provided that good starting values are
available.

\medskip

\noindent The drawback of working directly with
(\ref{direct2x}) is that an explicit expression for the
determinant of $H$ is required. To avoid this, let
$u,v\in\mathbb{C}^n$ be such that \be \label{eq:Hsig}
H\left(j\omega,\sigma,(\epsilon
w(\sigma))^{-1}\right) \left[\begin{array}{c}u\\
v\end{array}\right]=0,\ \ \ \hat{n}(u,v)=0, \ee where
$\hat{n}(u,v)=0$ is a normalizing condition. Given the
structure of $H$ it can be verified that a corresponding
left eigenvector is given by $[-v^*\ u^*]$. According to
\cite{Lancaster:99}, we get
\[
h_{\sigma}'(j\omega)=0\Leftrightarrow [-v^*\ u^*]\
\frac{\partial}{\partial
\lambda}H(j\omega,\sigma,(\epsilon
w(\sigma))^{-1}) \left[\begin{array}{c}u\\
v\end{array}\right]=0.
\]
A simple computation yields:
\begin{multline}
[-v^*\ u^*]\ \frac{\partial}{\partial
\lambda}H(j\omega,\sigma,(\epsilon
w(\sigma))^{-1}) \left[\begin{array}{c}u\\
v\end{array}\right]= \\
2\Im\left\{v^*\left(I+\sum_{i=1}^m A_{\sigma,i}\tau_i
e^{-j\omega\tau_i}\right)u\right\},
\end{multline}
which is always real. This is a consequence of the property
(\ref{col2}).

\medskip

\noindent Taking into account the above results, we end up with
$4n+3$ real equations
\begin{equation}\label{forfinal}
\left\{\begin{array}{l} H(j\omega,\ \sigma,\ (\epsilon
w(\sigma))^{-1})\left[\begin{array}{c}u, \\
v\end{array}\right]=0, \quad \hat{n}(u,v)=0\\
\Im\left\{v^*\left(I+\sum_{i=1}^m A_{\sigma,i}\tau_i e^{-j\omega\tau_i}\right)u\right\}=0\\
\end{array}\right.
\end{equation}
in  the $4n+2$ unknowns $\Re(v),\Im
(v),\Re(u),\Im(u),\omega$ and $\sigma$. These equations
are still overdetermined because  the property
(\ref{col1}) is not explicitly exploited in the
formulation, unlike the property (\ref{col2}). However,
it makes the equations (\ref{forfinal}) exactly solvable,
and the $(\omega,\sigma)$ components have a
one-to-one-correspondence with the solutions of
(\ref{direct2x}).

In our implementation the equations (\ref{forfinal}) are
solved using the Gauss-Newton method. This method exhibits
quadratic convergence because the residual in the
solution is zero, i.e., an exact solution exists \cite{BjorckBook}. The starting values are generated using
the approach outlined in the previous section.


\section{Algorithm} \label{sec:alg}

The overall algorithm for computing the \psa is as
follows.

\begin{alg} \label{alg:overall} \ \
${}$\\
\textit{Input:} system data, tolerance for prediction step, tol, and number of
discretization points, $N$\\
\textit{Output:} \psa $\alpha_\epsilon(F)$ \\

\underline{\textit{Prediction Step}:}
\begin{enumerate}
  \item[1)] Calculate the spectral abscissa $\alpha(F_N)$
    \item[2)] $\sigma_L=\alpha(F_N)$, $\sigma_R=\infty$, $\Delta \sigma =$tol,
  \item[3)] while $(\sigma_R-\sigma_L)>\textrm{tol}$
  \begin{enumerate}
  \item[3.1)] if $(\sigma_R=\infty)$ \\
              then $\Delta \sigma = 2 \times \Delta \sigma$, $\sigma_M=\sigma_L+\Delta \sigma$, \\
              else $\sigma_M=\frac{\sigma_L+\sigma_R}{2}$.
  \item[3.2)] if $H_{N,\sigma_M}$ has imaginary axis eigenvalues \\
              then $\sigma_L=\sigma_M$, \\
              else $\sigma_R=\sigma_M$.
  \end{enumerate}
   \item[] \hspace*{-0.7cm}\{result:
   $\alpha_{\epsilon}^N(F)=\sigma_L$ and
   $j\tilde \omega_i, i=1,\ldots,\tilde{n}$:  imaginary axis eigenvalues of $H_{N,{\sigma}_L}$\}
\end{enumerate}

\underline{\textit{Correction Step}:}
\begin{enumerate}
\item calculate the approximate null
vectors $\left\{x_1,\ldots,x_{\tilde{n}}\right\}$ of
$H(j\tilde{\w}_i,\alpha_\epsilon^N(F),(\epsilon
w(\alpha_\epsilon^N(F)))^{-1})$ $i=1,\ldots,\tilde{n}$,
 \item for all $i\in\{1,\ldots,\tilde{n}\}$, solve (\ref{forfinal}) with
 starting values
 \[
\left[\begin{array}{c}u\\ v\end{array}\right]=x_i,\
\omega=\tilde{\omega}_i,\ \ \sigma=\alpha_\epsilon^N(F)
 \]
denote the solution with $(u_{\epsilon,i},
v_{\epsilon,i},\omega_{\epsilon,i},\sigma_{\epsilon,i})$.
\item set
$\alpha_\epsilon(F):=\max_{1\leq i\leq \tilde{n}}
\sigma_{\epsilon,i}$.
\end{enumerate}
\end{alg}

The two steps are the prediction step explained in
Section~\ref{sec:psapred} and the correction step
explained in Section~\ref{sec:psacorr}. The first step
requires a \emph{repeated} computation of the eigenvalues
of the $2n(N+1)\times 2n(N+1)$ Hamiltonian matrix
$H_{N,\sigma}$ (\ref{eq:HamAN}). The second step solves
(\ref{forfinal}), i.e.\ a set of $4n+3$ nonlinear
equations. Our implementation chooses $N$ large enough
and the tolerance in the prediction step small enough
such that the results of the prediction step are good
starting values for the correction step.

Note that by increasing $N$ and reducing the tolerance, the
approximate \psa can be computed arbitrarily close to
$\alpha_\epsilon(F)$ by applying  the prediction step
only. However, this approach typically has a much larger
numerical cost than the combined approach, not only
because it requires a much larger  value of $N$ than
necessary for the corrector (to assure that $|\alpha_\epsilon(F)-\alpha_\epsilon^N(F)|$ sufficiently small), but also because the tolerance in the prediction step must be chosen very small (to assure that $\alpha_\epsilon^N(F)$ is computed sufficiently accurately). The latter implies that the number of iterations becomes very large. Hence, working with the prediction step only requires a much larger number of much more expensive iterations than working with the combined approach.

In our implementation, the mesh points in the
approximation of $F$, discussed in
Section~\ref{sec:approx}, are chosen as scaled and shifted Chebyshev
extremal points, i.e.,
\be \label{eq:chebpts}
\theta_{N,i}=\frac{\tau_{\max}}{2}\left(\cos\left(\frac{i\pi}{N}\right)-1\right),\ \ i=-N,\ldots,0
\ee since the corresponding interpolating
polynomial has less oscillation towards the end of the
interval compared to choices of grid points different from (\ref{eq:chebpts}), see \cite{Breda:06}.

Finally, we note that the prediction step is based on
approximating $F$ by $F_N$, defined in (\ref{eq:FN}),
hence, on approximating the exponential functions
$\lambda\mapsto \exp(-\lambda\tau_i)$ by the rational
functions $\lambda\mapsto p_N(-\tau_i;\ \lambda)$.
Because these approximations are essentially
approximations around $\lambda=0$, our implementation
incorporates the following substitution in $F(\lambda)$ to shift the center of the approximation to $\lambda=\alpha(F(\lambda))$:
\[
\lambda \leftarrow \lambda+\alpha(F(\lambda)),
\]
as well as a corresponding adaptation of the weights in the
pseudospectrum definition. For the computation of the
spectral abscissa $\alpha(F)$ we use the package
DDE-BIFTOOL,~\cite{Engelborghs:02}.

\section{Example} \label{sec:ex}

\begin{figure}
\begin{center}
\includegraphics[width=8cm]{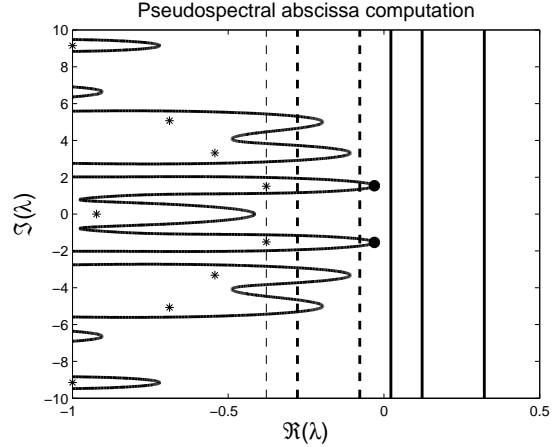}
\caption{\label{fig:example} The \ps and the pseudospectral abscissa. The stars indicate the characteristic roots of the time-delay system and the black curves are the pseudospectra contours. Vertical lines are lower and upper bounds in the bisection algorithm shown as dashed and solid lines respectively.}
\end{center}
\end{figure}

We tested the numerical method on several benchmark problems.
We chose the following high-order example with many delays to give further details about the algorithm.
We consider a time-delay system in (\ref{eq:DDEF}) with the dimensions
$m=7$, $n=10$ with delays $\tau_1=0.1$, $\tau_2=0.2$, $\tau_3=0.3$,
$\tau_4=0.4$, $\tau_5=0.5$, $\tau_6=0.6$, $\tau_7=0.8$. The weights $w_i$ are set to $1$
and $\epsilon=0.1$. The \ps is shown with black lines and black stars indicate part of the
characteristic roots of (\ref{eq:chareqn}) in Figure~\ref{fig:example}.

The tolerance in the bisection algorithm is set to $0.1$ and the discretization parameter is chosen as $N=10$.
Each iteration of the while loop in the prediction step computes $\sigma_M$ and updates $\sigma_L$ or $\sigma_R$ shown as the vertical dashed and solid lines respectively. The approximate \psa as a result of the prediction step is $\alpha_\epsilon^N(F)=-0.0525$ and the corresponding critical frequencies are $\tilde{\w}_1=1.4069$, $\tilde{\w}_2=1.6718$. These approximate values are improved in the correction
step and the computed \psa is
$\alpha_\epsilon(F)=-0.0307$ at $\w_\epsilon=1.5383$
shown as black dots in Figure~\ref{fig:example}.

In Table~\ref{table:Benchmarks} we present the results of
benchmarking of our code with $10$ time-delay plants with various perturbation sizes and perturbation weights.  The second
column shows the size of matrices $A_i$, $n$, and the
number of state delays, $m$. The third column gives the
minimum value of $N$ such that in the correction step the
desired solution is computed. The fourth and fifth
columns contain the predicted and corrected pseudospectral abscissa
of the corresponding time-delay system. The last column shows the computation time for each plant in seconds on a PC with an Intel Core Duo 2.53 GHz processor with 2 GB RAM. The plant $6$
corresponds to the problem considered in this section.

\begin{table}[h!]
\begin{center}
\begin{tabular}{l|l|l|l|l|l}
  Plants & $(n,m)$ & $N$ & $\alpha_\epsilon^N$ & $\alpha_\epsilon$ & time \\
  \hline
  \hline
  $1$ & $(3,1)$ & $6$ & $1.7784$ & $1.7790$ & $0.047$ \\
  $2$ & $(1,1)$ & $6$ & $4.1497$ & $4.1498$ & $0.048$\\
  $3$ & $(3,3)$ & $3$ & $5.5034$ & $5.5131$ & $0.061$\\
  $4$ & $(4,9)$ & $6$ & $6.4172$ & $6.4173$ & $0.075$\\
  $5$ & $(8,20)$ & $5$ & $7.2918$ & $7.2496$ & $0.27$\\
  $6$ & $(10,7)$ & $3$ & $-0.02840$ & $-0.03071$ & $0.19$\\
  $7$ & $(20,9)$ & $7$ & $3.4569$ & $3.4570$ & $3.05$\\
  $8$ & $(40,3)$ & $4$ & $1.2104$ & $1.2105$ & $4.66$\\
  $9^*$ & $(5,1)$ & $3$ & $1.9985$ & $1.9985$ & $0.10$\\
  $10^*$ & $(4,3)$ & $20$ & $1.5170$ & $1.5172$ & $0.60$\\
  \hline
\end{tabular}
\end{center}
\caption{Benchmarks for the pseudospectral abscissa computation.}
\label{table:Benchmarks}
\end{table}

For the plant $9$ a warning is generated when using the
default tolerance value of the prediction step $tol=10^{-3}$, indicating
that the difference between final lower and upper bound values for the approximate pseudospectral
abscissa is too large for the problem. The warning is removed when a smaller tolerance is chosen $tol=10^{-4}$. The plant $10$ gives a warning when the number of discretization points is set to the default value $N=15$. The warning is removed when $N=20$ is set. We note that both examples, plants $9$ and $10$, are difficult constructed cases. For most practical problems, the default values for the number of discretization points $N=15$ and the tolerance of the prediction step $tol=10^{-3}$ is sufficient.
\smallskip

The problem data for the above benchmark examples (system matrices $A_i$, state delays $\tau_i$, perturbation weights $w_i$ for $i=0,\ldots,m$, the perturbation size $\epsilon$ and options if necessary) and a MATLAB implementation of our code for the pseudospectral abscissa computation are available at the website
\begin{verbatim}
http://www.cs.kuleuven.be/~wimm/software/psa/
\end{verbatim}

\section{Concluding Remarks} \label{sec:conc}

An accurate method to compute the \psa of retarded
time-delay systems with an arbitrary number of delays is
given. The method is based on two steps: the prediction
step calculates an approximation of the \psa  based on a
finite-dimensional approximation of the problem. The
correction step computes the \psa by solving nonlinear
equations that characterize the rightmost points of the
pseudospectrum. The method has been successfully applied to
benchmark problems demonstrating its effectiveness.

After the \psa of the time-delay plant is computed, the gradient of the \psa with respect to system matrices and delays can be calculated for the complex point where \psa is achieved. By embedding the pseudospectral abscissa computation in an optimization loop, a fixed structure controller minimizing the \psa can be designed inspired by the approach of \cite{HIFOO:Hafia} for the finite dimensional case. This is our future research direction.

\begin{ack}
This article present results of the Belgian Programme on
Interuniversity Poles of Attraction, initiated by the Belgian State,
Prime Minister's Office for Science, Technology and Culture, and of
OPTEC, the Optimization in Engineering Centre of the K.U.Leuven.
\end{ack}

\bibliographystyle{plain}        
\bibliography{journal_psa_rev}

\appendix
\section{Proof of Theorem~\ref{maintheo}} \label{sec:app}
We need the following proposition to prove the Theorem
\ref{maintheo}.
\begin{proposition}\label{prop1}
We can express
\begin{equation}\label{lem1}
p_N(-\tau_i;\ \lambda)=\frac{r_i(\lambda)}{s(\lambda)},\
1=1,\ldots,m,
\end{equation}
where $s$ is a monic polynomial of degree $N$ and $r_i,\
i=1,\ldots,m$\ are polynomials of degree smaller than or
equal to $N$. Furthermore, we have
\begin{equation}\label{lem2}
s(\lambda)=\det(\lambda I-D_{1,1})
\end{equation}
and
\begin{equation}\label{lem3}
\left[\begin{array}{c} p_N(\theta_{N,-N};\ \lambda)\\
\vdots\\
p_N(\theta_{N,-1};\ \lambda)
\end{array}\right]=(\lambda I-D_{1,1})^{-1}D_{1,2}.
\end{equation}
where $D_{1,1}$ and $D_{1,2}$ are given in (\ref{eq:D}).
\end{proposition}

\noindent\textbf{Proof.\ } In a Lagrange basis we can
express
\[
p_N(t;\ \lambda)=\sum_{i=-N}^0 c_i l_{N,i}(t),
\]
where, for the simplicity of the notations, we suppress
the dependence of the coefficients $c_i$ on $\lambda$.
The conditions (\ref{defpn}) can be expressed as $c_0=1$
and
\[
(\lambda I-D_{1,1})\left[\begin{array}{l}c_{-N} \\ \vdots
\\ c_{-1}\end{array}\right]=D_{1,2},
\]
which implies that
\[
p_N(t;\ \lambda)=l_{N,0}(t)+[l_{N,-N}(t)\ \cdots\
l_{N,-1}] (\lambda I-D_{1,1})^{-1} D_{1,2}.
\]
The assertions follow. \hfill$\Box$

\noindent\textbf{Proof of Theorem \ref{maintheo}.\ }
Using the formula for the determinant of a two-by-two
block matrix based on Schur complements and with ${\bf A_N}$ and $D$ given in (\ref{eq:AN}) and (\ref{eq:D}) respectively, it
follows that
\begin{multline}
\nonumber \det(\lambda I-{\bf A_N})=\det((\lambda I_N-D_{1,1})\otimes I_n) \det(\lambda I_n - \Gamma_0 \\
-[\Gamma_{-N}\cdots \Gamma_{-1}] ((\lambda I_N-D_{1,1})\otimes I_n)^{-1} (D_{1,2}\otimes I_n) ),
\end{multline}\vspace{-.7cm}
\begin{displaymath}
=s(\lambda)^n\det\left(\lambda I_n-\Gamma_0-\sum_{i=-N}^{-1}\Gamma_i I_n \ p_N(\theta_{N,i};\ \lambda ) \right),
\end{displaymath}\vspace{-.7cm}
\begin{multline}
\nonumber \hspace{-1mm} =s(\lambda)^n\det\left(\lambda I_n-A_0 \right.\\
\left.-\sum_{i=-N}^{0}\sum_{l=1}^m A_l l_{N,i} (-\tau_l)p_N(\theta_{N,i};\ \lambda ) \right),
\end{multline}\vspace{-.7cm}
\begin{multline}
\nonumber \hspace{-1mm}=s(\lambda)^n\det\left(\lambda I_n-A_0-\sum_{l=1}^m  A_l \right.\\
\left.\sum_{i=-N}^{0}l_{N,i} (-\tau_l) p_N(\theta_{N,i};\ \lambda ) \right),
\end{multline}\vspace{-.4cm}
\begin{equation}\label{approx1}
=s(\lambda)^n\det\left(\lambda  I_n-
A_0-\sum_{l=1}^m  A_l p_N(-\tau_l;\ \lambda ) \right).
\end{equation}

Furthermore, using the same approach, we can derive for $k,l\in\{1,\ldots,n\}$:
\begin{multline} \label{approx2}
\Delta^{k,l}_N(\lambda):=\left\{{\bf B_N}^*\
\mathrm{adj}(\lambda I_{(N+1)n}-{\bf A_N}){\bf B_N}\right\}_{k,l},
\\
=\det((\lambda I_N-D_{1,1})\otimes I_n) \det(\lambda
\tilde I_{n-1}-\tilde \Gamma_0-[\tilde\Gamma_{-N}\cdots
\tilde\Gamma_{-1}] \\
((\lambda I_N-D_{1,1})\otimes I_n)^{-1}
(D_{1,2}\otimes \tilde I_n)),
\end{multline}
where the superscript \verb|~|  denotes that an
appropriate row and/or column have been removed. Using
Proposition~\ref{prop1} and following the steps in
(\ref{approx1}), this expression can be written as

\begin{equation}\label{approx3}
\begin{array}{l}
\Delta_N^{k,l}(\lambda) =s(\lambda)^n\det\left(\lambda \tilde I-\tilde \Gamma_0-\sum_{i=-N}^{-1}\tilde\Gamma_i \tilde I \ p_N(\theta_{N,i};\ \lambda ) \right) \\
=s(\lambda)^n \left\{ \mathrm{adj}\left( \lambda I-
A_0-\sum_{l=1}^m  A_l
 \ p_N(-\tau_l;\ \lambda )
\right) \right\}_{k,l}.
\end{array}
\end{equation}
Using (\ref{approx1})-(\ref{approx3}) we can derive:
\[
\begin{array}{l}
{\bf B_N}^T(\lambda I-{\bf A_N})^{-1} {\bf B_N}
= {\bf B_N}^T \frac{\mathrm{adj}(\lambda I-{\bf A_N})}{\det(\lambda I-{\bf A_N})} {\bf B_N}\\
=\frac{\mathrm{adj}\left(\lambda I-A_0-\sum_{i=1}^m A_i
p_N(-\tau_i;\ \lambda)\right) }{\det\left(\lambda
I-A_0-\sum_{i=1}^m A_i p_N(-\tau_i;\ \lambda)\right)}
\\
=(\lambda I-A_0-\sum_{i=1}^m A_i p_N(-\tau_i;\
\lambda))^{-1}.
\end{array}
\]
This completes the proof. \hfill $\Box$

\end{document}